\documentclass[twoside,twocolumn,english,aps,prl,longbibliography,floatfix,superscriptaddress]{revtex4-1}
\usepackage{color}
\usepackage{textcomp}
\usepackage{amsmath}
\usepackage{amssymb}
\usepackage{graphicx}
\usepackage{bbold}
\usepackage{natbib}
\usepackage{makecell}
\usepackage{booktabs}
\usepackage{esint}
\usepackage{dcolumn}
\usepackage{bm}
\usepackage[normalem]{ulem}
\usepackage{makecell, multirow}
\usepackage{subfloat}
\usepackage[table]{xcolor}
\usepackage[caption=false,singlelinecheck=off]{subfig}
\makeatletter
\@ifundefined{textcolor}{}
{%
 \definecolor{BLACK}{gray}{0}
 \definecolor{WHITE}{gray}{1}
 \definecolor{RED}{rgb}{1,0,0}
 \definecolor{GREEN}{rgb}{0,1,0}
 \definecolor{BLUE}{rgb}{0,0,1}
 \definecolor{CYAN}{cmyk}{1,0,0,0}
 \definecolor{MAGENTA}{cmyk}{0,1,0,0}
 \definecolor{YELLOW}{cmyk}{0,0,1,0}
}
\usepackage[colorlinks=true,linkcolor=blue,urlcolor=blue,citecolor=blue]{hyperref} 
\hypersetup{colorlinks=true}
\usepackage[all]{hypcap} 

\begin{document}
\title{Noise on the Non-Abelian $\nu=5/2$ Fractional Quantum Hall Edge}
\author{Jinhong Park}
\email{jinhong@thp.uni-koeln.de}
\affiliation{Institute for Theoretical Physics, University of Cologne, Z\"{u}lpicher Str. 77, 50937 K\"{o}ln, Germany}
\affiliation{Department of Condensed Matter Physics, Weizmann Institute of Science, Rehovot 76100, Israel}
\author{Christian Sp\r{a}nsl\"{a}tt}
\email{christian.spanslatt@kit.edu}
\affiliation{Institute for Quantum Materials and Technologies, Karlsruhe Institute of Technology, 76021 Karlsruhe, Germany}
\affiliation{Institut f\"{u}r Theorie der Kondensierten Materie, Karlsruhe Institute of Technology, 76128 Karlsruhe, Germany}
\author{Yuval Gefen}
\affiliation{Department of Condensed Matter Physics, Weizmann Institute of Science, Rehovot 76100, Israel}
\affiliation{Institute for Quantum Materials and Technologies, Karlsruhe Institute of Technology, 76021 Karlsruhe, Germany}
\author{Alexander D. Mirlin}
\affiliation{Institute for Quantum Materials and Technologies, Karlsruhe Institute of Technology, 76021 Karlsruhe, Germany}
\affiliation{Institut f\"{u}r Theorie der Kondensierten Materie, Karlsruhe Institute of Technology, 76128 Karlsruhe, Germany}
\affiliation{Petersburg Nuclear Physics Institute, 188300 St. Petersburg, Russia}
\affiliation{L.\,D.~Landau Institute for Theoretical Physics RAS, 119334 Moscow, Russia}

\date{\today}
\begin{abstract}
The recent measurement of a half-integer thermal conductance for the $\nu=5/2$ fractional quantum Hall state has confirmed its non-Abelian nature, making the question of the underlying topological order highly intriguing. We analyze the shot noise at the edge of the three most prominent non-Abelian candidate states. We show that the noise scaling with respect to the edge length can, in combination with the thermal conductance, be used to experimentally distinguish between the Pfaffian, anti-Pfaffian, and particle-hole-Pfaffian edge structures.
\end{abstract}
\maketitle

\textit{Introduction.---} 
The fractional quantum Hall (FQH)~\cite{Stormer1982,Laughlin1983} state at filling $\nu=5/2$~\cite{Willet1987} is the prototypical candidate for a phase of matter with non-Abelian topological order~\cite{Moore1991}. Such order has attracted immense attention during the last decades, not least for its remarkably rich theoretical structure~\cite{Fradkin2013Book}, but also as a promising platform for topological quantum computation~\cite{Nayak2008}. 

\begin{figure*}[t!]
\label{fig:Fig1}
\captionsetup[subfigure]{position=top,justification=raggedright}
\begin{tabular}{l}
	\subfloat[]{
\includegraphics[width=0.3\textwidth]{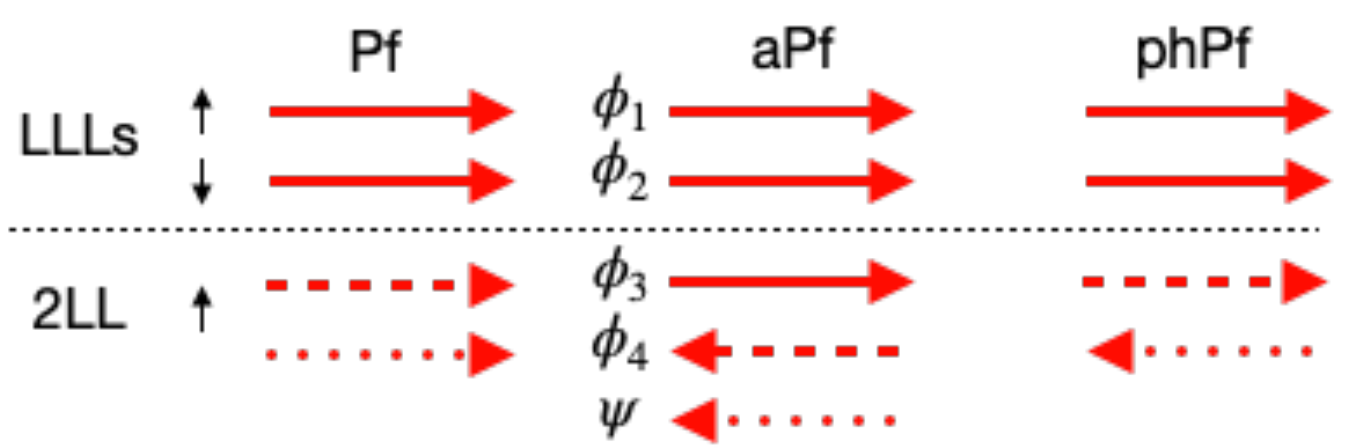}
\label{fig:LLs}} \\
\subfloat[]{
\includegraphics[width=0.3\textwidth]{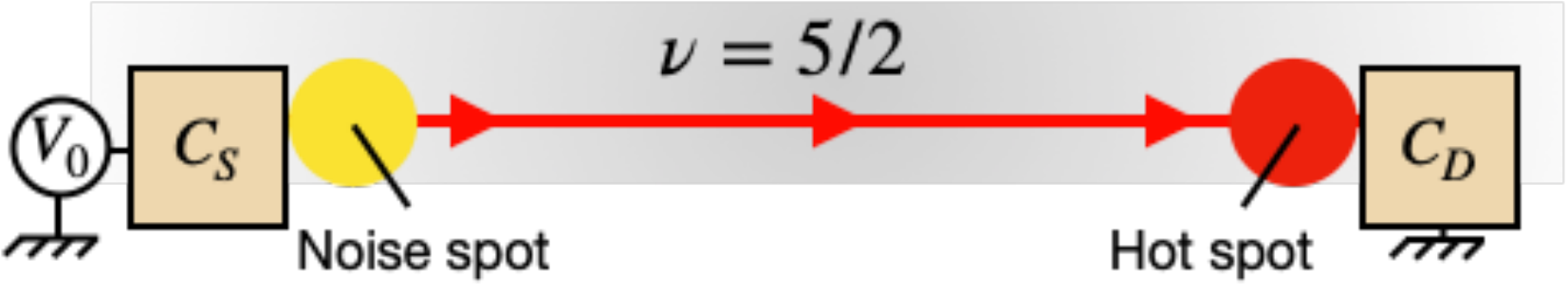}
\label{fig:Setup}}
\end{tabular}
\hspace*{-0.3cm}
\begin{tabular}{c}
	\subfloat[]{
\includegraphics[width=0.36\textwidth]{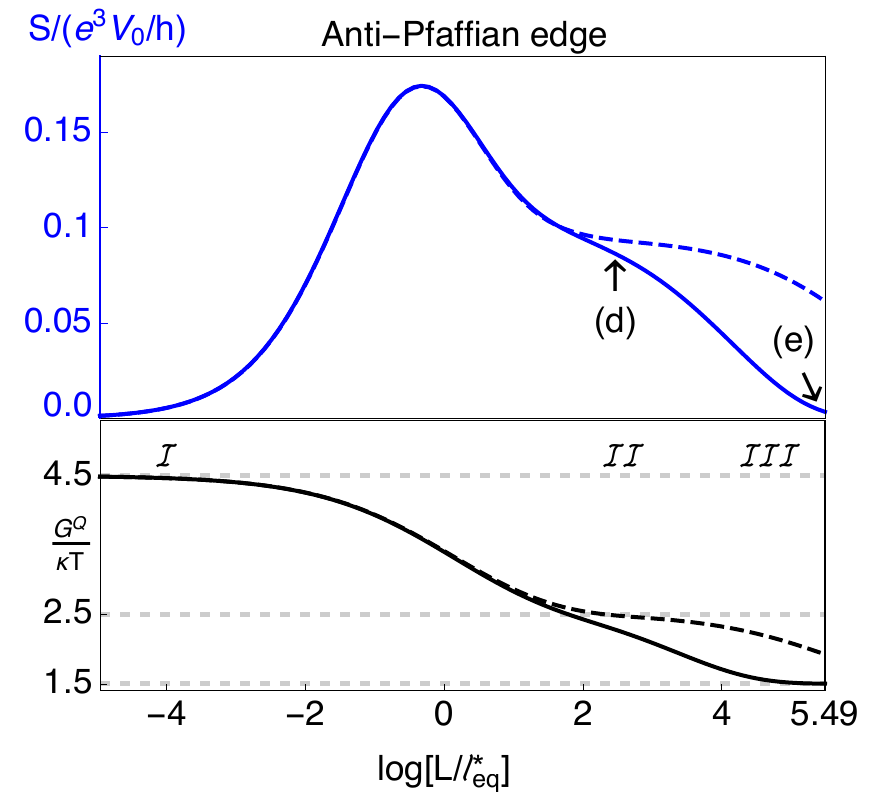}
\label{fig:Noise}}
\end{tabular}
\hspace*{-0.3cm}
\begin{tabular}{r}
\subfloat[]{
\includegraphics[width=0.29\textwidth]{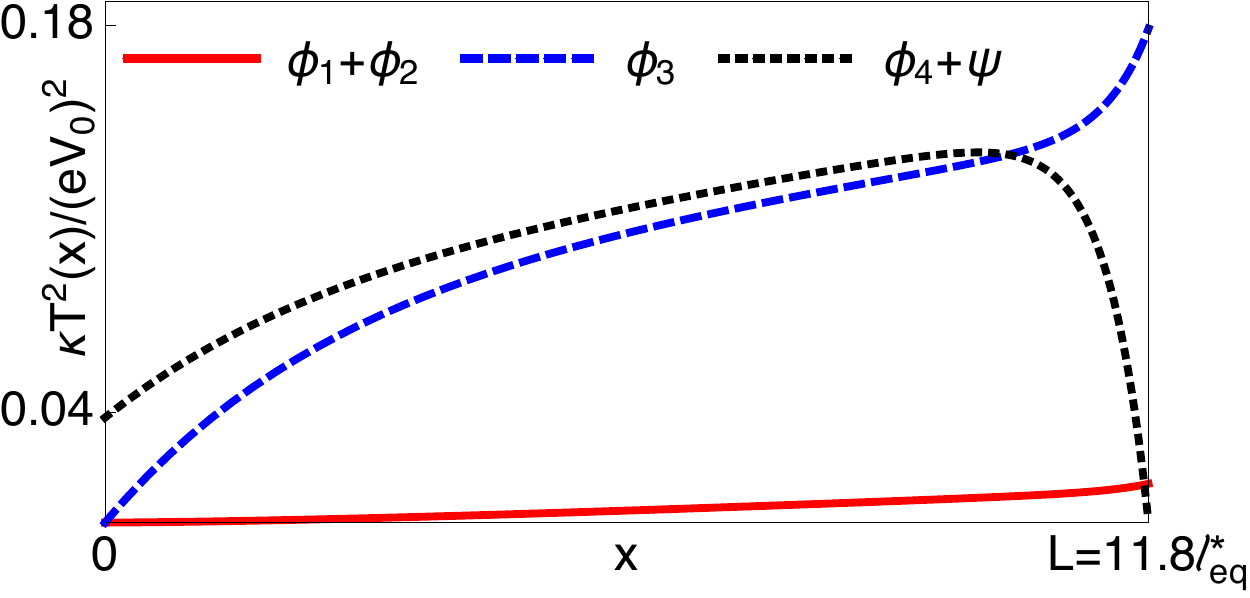}
\label{fig:TAB}}\\
\subfloat[]{
\includegraphics[width=0.29\textwidth]{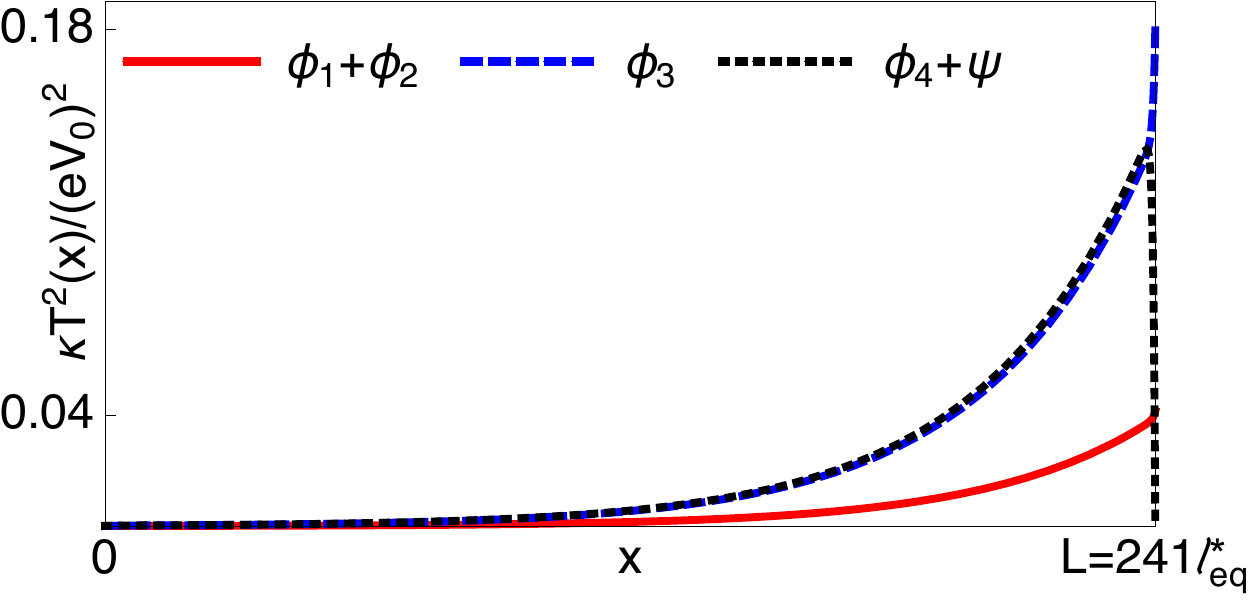}
\label{fig:TB}}
\end{tabular}
\caption{\label{fig:NoiseAndSetup} (a) Lowest (LLL) and second Landau level (2LL) edge structures of Pfaffian (Pf), anti-Pfaffian (aPf), and particle-hole Pfaffian (phPf) states. 
Thick lines: unit-charge bosonic channels, dashed lines: charge $1/2$ bosons, dotted lines: Majorana channels. Arrows denote downstream (right-pointing) or upstream (left-pointing) propagation. Black arrows indicate spin.
 (b) Schematic setup for measurement of noise $S$. The contacts are separated by distance $L$, one of them is biased with $V_0$.  In the equilibrated regime,
 heat is generated at the hot spot (red dot) at the downstream contact, while noise is generated due to partitioning of electron-hole pairs at the noise spot (yellow dot). The noise is independent of the direction of the applied bias. (c) Shot noise $S/(V_0e^3/h)$ and thermal conductance $G^Q/\kappa T$ of the aPf edge as functions of $\text{log}[L/\ell_{\rm eq}^*]$ for $\ell_{\rm eq}=100$ (solid lines) and $\ell_{\rm eq}=1000$ (dashed lines). In regime $\mathcal{I}$ (see Tab.~\ref{tab:TransportTab}) the equilibration and $S$ are weak and $G^Q/\kappa T=9/2$. In regime $\mathcal{II}$ where LLM is weak but intra-2LL equilibration efficient, $S$ is approximately constant and $G^Q/\kappa T\approx 5/2$. In regime $\mathcal{III}$ with full equilibration, $S$ is exponentially suppressed and $G^Q/\kappa T\rightarrow 3/2$. (d) aPf edge channel temperature profiles in regime $\mathcal{II}$ with $\ell_{\rm eq}/\ell^*_{\rm eq}=100$ and $L\approx 11.8 \ell^*_{\rm eq}$. Heat from the hot spot ($L-\ell_{\textrm{eq}}^*\lesssim x\lesssim L$) reaches the noise spot ($0\lesssim x\lesssim \ell_{\textrm{eq}}^*$). (e) aPf edge channel temperature profiles in regime $\mathcal{III}$ with $\ell_{\rm eq}/\ell^*_{\rm eq}=100$ and $L\approx 241 \ell^*_{\rm eq}$. The heat reaching the noise spot is exponentially small in $L$.}
\end{figure*}

The $5/2$ state is believed to consist of two filled lowest Landau levels (LLLs) with opposite spin-polarizations and one half-filled and spin polarized second Landau level (2LL)~\cite{Morf1998,Park1998,Feiguin2009,Pan2001,Tiemann2012,Stern2012,Biddle2013}. For this structure, a wide variety of theoretical candidate states have been proposed, among which the most prominent are the Pfaffian (Pf)~\cite{Moore1991}, anti-Pfaffian (aPf)~\cite{Levin2007,Lee2007} and particle-hole Pfaffian (phPf)~\cite{Fidkowski2013,Son2015,Zucker2016,Antonic2018} states, which all exhibit non-Abelian order. Also several Abelian states have been proposed~\cite{Wen1991,Halperin1983,Yang2013,Yang2014}. To date, numerical simulations seem to favor the aPf state~\cite{Morf1998,Storni2010,Rezayi2017}, while tunneling experiments point either towards the aPf, $SU(2)_2$, $331$, or $113$ states~\cite{Radu2008,Lin2012,Lin2014}. All proposed candidates are compatible with the Hall conductance $G_H = 5e^2/2h$, but they differ in their bulk topological order, manifested by different edge structures~\cite{Wen1990a,Wen1990b,Chang2003} (see Fig.~\ref{fig:LLs}). A fruitful route in determining the nature of the $5/2$ state is therefore by thermal edge transport experiments~\cite{Jezouin2013,Banerjee2017,Banerjee2018,Heiblum2019}. If the edge fully equilibrates due to efficient inter-channel tunneling, the thermal Hall $G^Q_H$ and two-terminal $G^Q$ conductances are quantized as
\begin{equation}
\label{eq:GQ}
G^Q_H = \nu_Q\kappa T, \qquad G^Q = |G^Q_H|,
\end{equation}
where, $\kappa = \pi^2 k_B^2/3h$, $T$ is the temperature, and $k_B$ is Boltzmann's constant. The topological quantity $\nu_Q\equiv c-\bar{c}$ is the difference in the central charges of the chiral ($c$) and the anti-chiral ($\bar{c}$) sectors of the edge conformal field theory~\cite{Kane1997,Capelli2002}. It should be emphasized however that, with insufficient equilibration, $G^Q/\kappa T$ may in principle take any value between $c+\overline{c}$ and $|\nu_Q|$. For an Abelian edge, $c$ and $\bar{c}$ coincide with the number of downstream (the chirality direction set by the magnetic field) and upstream (opposite direction to downstream) edge channels respectively~\cite{Kane1997,Capelli2002}. By contrast, a chiral Majorana edge mode $\psi$, present only on non-Abelian edges, contributes instead with $c_{\rm \psi}=1/2$, implying a half-integer quantization in Eq.~\eqref{eq:GQ}. Indeed, Banerjee \textit{et al.}~\cite{Banerjee2018} recently found $G^Q/\kappa T \approx 5/2$; a clear signature of non-Abelian order. This particular value of $G^Q$ was further interpreted as favoring the phPf state for which $\nu_Q=5/2$. Under certain conditions, this particle-hole symmetric value of $\nu_Q$ can also be obtained in models with random puddles of alternating non-Abelian orders~\cite{Mross2018,Wang2018,Biao2018,Zhu2020}. At the same time, theories of partial equilibration have been put forward, allowing the aPf edge to remain a viable candidate~\cite{Simon2018,Feldman2018,Ma2019,Simon2020,Asasi2020}. To our knowledge, no reconciliation between experiment and theory for the pure Pf edge, where $G^Q_H/\kappa T=7/2$ regardless of equilibration, has so far been made. Hence, the question whether the $\nu=5/2$ state displays aPf or phPf topological order remains open and pressing.

In this Letter, we propose that shot noise~\cite{Blanter2000} measurements are a powerful tool to distinguish between all three non-Abelian $5/2$ candidate states (see Fig.~\ref{fig:Setup}). We show that in the transport regime with complete edge equilibration, which requires strong Landau level mixing (LLM), the dc noice $S$ either vanishes or decreases exponentially with increasing edge length $L$. However, in the transport regime where LLM is negligible but equilibration within the 2LL is efficient, the aPf edge uniquely exhibits the scaling $S\simeq c_1 - c_2 \sqrt{L/\ell_{\textrm{eq}}}$ with constants $c_1, c_2 > 0$ (see Fig.~\ref{fig:Noise}). Most interestingly, it is precisely in this semi-equilibrated regime that $G^Q/\kappa T=5/2$ for both the aPf and phPf states. It follows that in combination with measurements of $G^Q$, the scaling of $S$ with $L$ uniquely distinguishes between the aPf and phPf edges. 

These disparate scalings follow from a delicate interplay of charge and heat transport on the FQH edge. With strong equilibration, $L/\ell_{\rm eq}\gg 1$, where $\ell_{\rm eq}$ is a characteristic length~\cite{GammaNote} for charge and heat equilibration~\cite{Kane1995,Protopopov2017,Nosiglia2018,Aharon2019}, noise is generated due to thermal partitioning of the charge current by the following mechanism~\cite{Park2019,Spanslatt2019,Spanslatt2020}. It is a remarkable consequence of the chiral edge nature that, when a current is driven between two contacts along an equilibrated FQH edge, heat is generated near the downstream contact (the hot spot), while noise near the upstream contact (the noise spot) (see Fig.~\ref{fig:Setup}). Thus, noise generation requires a heat flow {\it upstream} from the hot spot to the noise spot,  implying a deep connection between the noise characteristics and the nature of the heat transport along the edge. Since the latter is inherited from the bulk  topological order, the topological significance of the noise scaling follows. For edges with $\nu_Q>0$, $S\simeq 0$ (up to exponential corrections in $L/\ell_{\rm eq}$); for $\nu_Q=0$, $S\simeq \sqrt{\ell_{\rm eq}/L}$ and for $\nu_Q<0$, $S\simeq  {\rm const}$. Hence, this noise classification constitutes a powerful probe for the FQH edge structure and provides a fully electrical method to detect upstream heat propagation.

To apply this classification to the three non-Abelian $\nu=5/2$ edge candidates, we first define for each edge two length scales $\ell_{\rm eq}^*$ and $\ell_{\rm eq}$, which characterize intra-2LL and complete equilibration, respectively~\cite{LLLEquili}. We assume $\ell_{\rm eq}^* \ll \ell_{\rm eq}$, which will be justified below. Next, we identify transport coefficients and noise scaling for the candidate edges in three transport regimes: $L \ll \ell^*_{\rm eq}$ (regime $\mathcal{I}$, clean regime), $\ell_{\rm eq}^*\ll L \ll \ell_{\rm eq}$ ($\mathcal{II}$, no LLM), and $\ell_{\rm eq}\ll L$ ($\mathcal{III}$, full equilibration) [see Tab.~\ref{tab:TransportTab}]. For the maximally chiral Pf edge, no backscattering of charge or heat occurs. Thus, there is no charge partitioning and the noise vanishes identically in all regimes.  For the phPf edge, charge propagates only downstream as well,  hence no partitioning of the current and vanishing noise in all regimes~\cite{note-edge-reconstruction}. These results are to be contrasted with the aPf edge, which has a richer edge structure and is in the focus of this work. In regime $\mathcal{I}$, we assume $S\propto L$ due to rare scattering events (see Ref.~\onlinecite{Park2019} for details). In regime $\mathcal{III}$, Eq.~\eqref{eq:GQ} gives $\nu_Q = 3/2$, which by our classification implies exponentially suppressed $S$. However, when the LLM is weak, i.e., in regime $\mathcal{II}$, most of the noise is generated only in the 2LL due to a lack of backscattering in the two LLLs (which are to a large extent decoupled from the 2LL). The 2LL channels, within which heat flow upstream since $(c-\overline{c})|_{2LL}=-1/2$, lead to a constant noise $S\simeq c_1 - c_2 \sqrt{L/\ell_{\textrm{eq}}}$ up to algebraic correction in $L$. This algebraic correction originates from the weak heat loss of the 2LLs to two LLLs. The existence of this noisy regime for the aPf edge is our central observation. To investigate this regime,  we next perform a detailed  renormalization group (RG) analysis~\cite{Protopopov2017,Park2020} of $\ell_{\rm eq}^*$ and $\ell_{\rm eq}$ for the aPf edge.

\textit{Analysis of equilibration on the aPf edge.---}
The aPf edge consists of one left-moving charge neutral Majorana channel $\psi$  (with velocity $v_n$) and four charged bosonic channels $\phi_i$ ($i=1,\hdots,4$), where $\phi_4$ is left-moving while the others are rightmovers~\cite{Levin2007,Lee2007} (see Fig.~\ref{fig:LLs}). The action is $S=S_0+S_\psi$, with
\begin{align}
\label{eq:FullAction}
&S_0 = -\int dt dx \sum_{ij} \frac{1}{4\pi} \big[ K_{ij} \partial_x \phi_i \partial_t \phi_j + V_{ij} \partial_x \phi_i \partial_x \phi_j \big], \nonumber \\
&S_\psi = \int dt dx\big[ i \psi ( \partial_t - v_n \partial_x ) \psi \big].
\end{align} 
Here, the topological matrix $K = \rm{diag} (1, 1, 1, -2)$ in the basis $(\phi_1, \phi_2, \phi_3, \phi_4)$, and the non-universal matrix $V$ contains on its diagonal all bosonic velocities, while the off-diagonal elements describe inter-channel repulsive interactions. We ignore density-density interactions involving the Majorana, since these are RG irrelevant at low temperatures. The action~\eqref{eq:FullAction} is integrable and involves no mechanism for equilibration between the channels. In the absence of such a mechanism, we have $G/(e^2/h) = \sum_i|K^{-1}_{ii}|=7/2$ and $G^Q/\kappa T = 4+1/2=9/2$. We can introduce equilibration by adding random inter-channel electron tunneling~\cite{Kane1994}. Assuming that channels in the 2LL are spatially far away from the LLL channels, equilibration occurs dominantly within  the 2LL (see e.g., Ref.~\onlinecite{Ma2019}). We may then add the following random disorder perturbation~\cite{Levin2007}
\begin{equation} \label{eq:disorderAction}
	S_{\rm 2LL} =  \int dt dx \left [\xi_{\rm{2LL}}(x) \psi e^{i2\phi_4 + i\phi_3} + \rm{H.c.} \right],
\end{equation}
where $e^{i\phi_3}$ annihilates a right-moving electron while $\psi e^{i2\phi_4}$ creates a left-moving electron. For simplicity, we take $\xi_{\rm{2LL}}(x)$ as a complex Gaussian random variable, $\langle \xi_{\rm{2LL}}^{} (x)\xi_{\rm{2LL}}^*(x')\rangle =W_{\rm{2LL}} \delta(x-x')$.  

We now analyze the influence of this disorder on the edge transport by considering the linear RG equation for $W_{\rm 2LL}$. From the standard disordered averaged RG scheme~\cite{Giamarchi1988} we have $d\tilde{W}_{\rm{2LL}} / d\ln\ell = (3- 2\Delta_{\rm{2LL}}) \tilde{W}_{\rm{2LL}}$. Here, $\ell$ denotes the running length scale, $\Delta_{\rm{2LL}}$ is the scaling dimension of $\psi e^{i2\phi_4 + i\phi_3}$, and $\tilde{W}_{\rm{2LL}}$ is the dimensionless disorder strength corresponding to $W_{\rm{2LL}}$. Hereafter, all appearing dimensionless disorder strengths are denoted with tilde.
When the perturbation~\eqref{eq:disorderAction} is relevant ($\Delta_{\rm{2LL}} < 3/2$), the disorder drives the system towards the fixed point $\Delta_{\rm{2LL}} = 1$~\cite{Levin2007}. The RG flow then introduces an elastic length scale $\ell_0$ beyond which disorder mixes the channels within the 2LL. We define $\ell_0$ as the scale at which $\tilde{W}_{\rm{2LL}}$ is of order unity: $\ell_0 \sim a \tilde{W}_{\rm{2LL}, 0}^{1/(3-2\Delta_{\rm{2LL}})}$, where $a$ is the UV length cutoff  and $\tilde{W}_{\rm{2LL}, 0}\equiv\tilde{W}_{\rm{2LL}}(\ell =a)$. If the edge length $L$ is larger than $\ell_0$, the system flows towards the fixed point where it finally decouples into three upstream-propagating neutral Majorana modes $\psi_{a}$ ($a = 1,2,3$) and three downstream-propagating charge bosonic modes $\phi_1$, $\phi_2$, and $\phi_{\rho}=\phi_{3}+\phi_4$~\cite{Levin2007,Lee2007}. In the vicinity of this fixed point, $\ell_0$ constitutes the new UV cutoff for the RG analysis below. 

\begin{table}[t!]
\begin{ruledtabular}
\begin{tabular}{ccccc}  
& Transport characteristics & Pf & aPf & phPf  \\
\hline
\multirow{3}{*}{$\mathcal{I}$}   &  $G / (e^2 / h)$ &5/2 & 7/2 & 5/2  \\ 
 & $G^Q / (\kappa T)$& 7/2 & 9/2 & 7/2 \\
  &  $S$ & $0$& $\propto L$ & $0$\\
 \hline 
 \multirow{3}{*}{$\mathcal{II}$}   &  $G / (e^2 / h)$ & 5/2 & 5/2 & 5/2\\
\cline{4-5}
    &  $G^Q / (\kappa T)$ & 7/2 &  \multicolumn{1}{|c}{5/2} & \multicolumn{1}{c|}{5/2}\\
     &  $S$ & $0$ & \multicolumn{1}{|c}{\rm{const.}} & \multicolumn{1}{c|}{0}\\
\hline
 \multirow{3}{*}{$\mathcal{III}$} &$G / (e^2 / h)$ &5/2  & 5/2 & 5/2 \\
 &  $G^Q / (\kappa T)$ &7/2 & 3/2 & 5/2\\
& $S$ & $0$ & $\simeq$ 0 &  0
\end{tabular}
\end{ruledtabular}
\caption{\label{tab:TransportTab} Two-terminal electrical ($G$) and thermal ($G^{Q}$) conductances, and scaling of shot noise ($S$) with length $L$  for Pfaffian (Pf), anti-Pfaffian (aPf), and particle-hole Pfaffian (phPf) edges. Regime $\mathcal{I}$:  no equilibration,  regime $\mathcal{II}$: complete 2LL equilibration,  regime $\mathcal{III}$: full equilibration. $S\simeq 0$ means exponentially small noise $S \sim e^{-L/\ell^*_{\rm eq}}$. The marked box, where aPf and phPf edges show distinct noise scaling for the same $G^Q$, is a central result of this paper.}
\end{table}

\begin{figure}[t!]
\label{fig:PhaseDiagram}
\includegraphics[width=0.9\columnwidth]{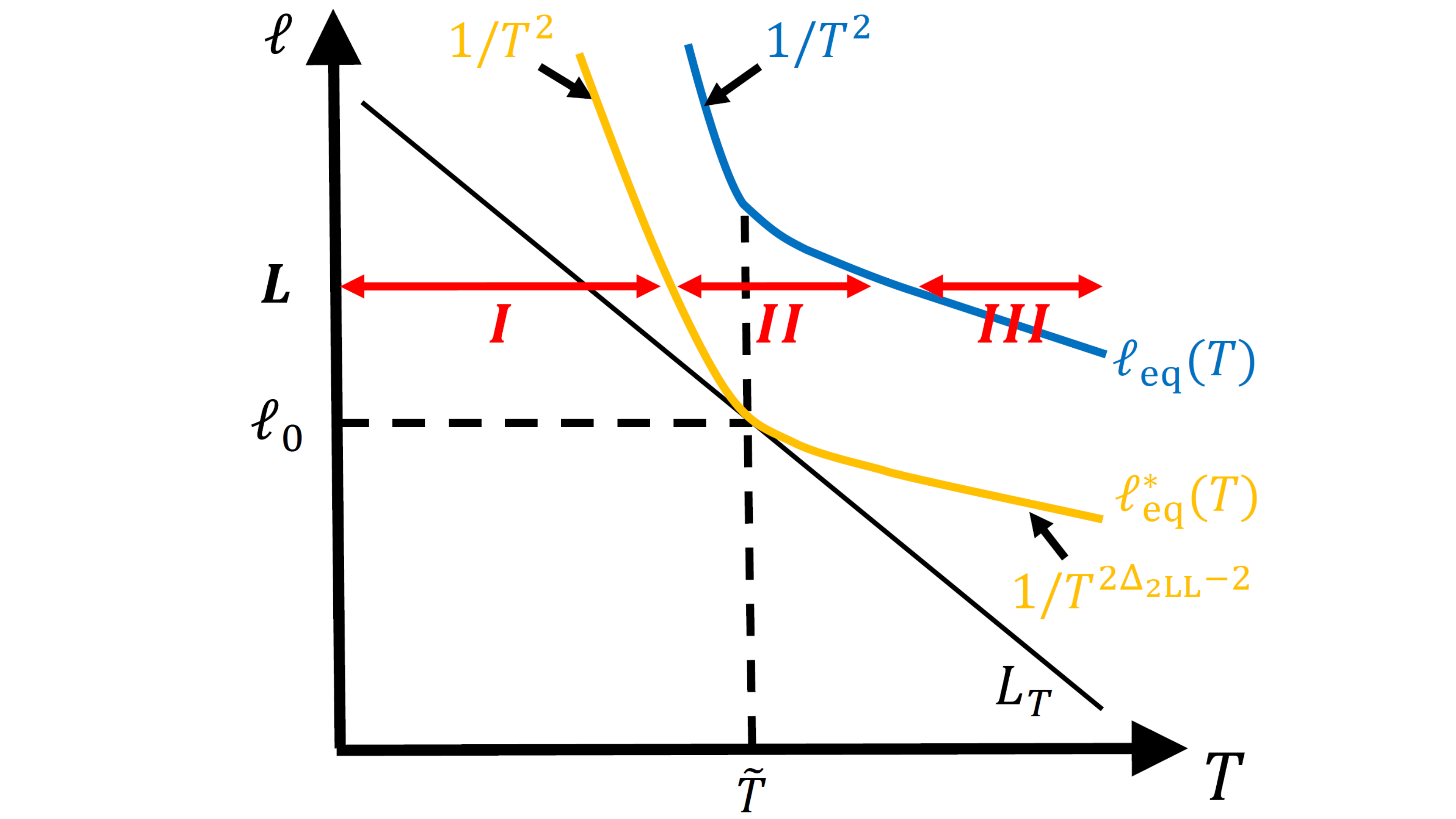}
\caption{Schematic log-log plot of the temperature ($T$) dependence of 
equilibration lengths $\ell^*_{\rm eq}$ (within the 2LL)
and $\ell_{\rm eq}$ (between the LLLs and the 2LL) for strong interactions $\Delta_{2LL}<3/2$. $\ell_{0}$ is a $T$-independent elastic length beyond which channels in the 2LL mix by disorder and the system enters the disorder-dominated phase, while $L_{T} \propto 1/T$ (black thin line) is the thermal length. The scaling of $\ell^*_{\rm eq}$ and $\ell_{\rm eq}$ changes at $T = \tilde{T}$, where the transition temperature $\tilde{T}$ is defined as $ L_T (\tilde{T}) = \ell_{0}$. For a given edge length $L$, three transport regimes $\mathcal{I}$, $\mathcal{II}$, and $\mathcal{III}$ are indicated (see Tab.~\ref{tab:TransportTab}). $T$ is replaced by the voltage $V$ when $k_B T \ll eV$.  }
\end{figure}

We then consider the length $\ell_{\rm{eq}}^{*}$ and its scaling with $T$, assuming $k_B T \gg eV$, where $V$ is the voltage bias. We first consider sufficiently low temperature ($T < \tilde{T}$ in Fig.~\ref{fig:PhaseDiagram}). In the vicinity of the fixed point, and in the basis of charged bosons and neutral Majoranas, the part of $S+S_{\rm 2LL}$ equilibrating the 2LL reads
\begin{equation}
 S_{\psi \rho} = -  \frac{v_{\rho \sigma}}{2\pi} \sum_{a \neq b} \int dx dt  \partial_x \phi_\rho \psi_a  (R^T(x) L_x R(x))_{ab} \psi_b. 
 \nonumber
\end{equation}
Here, $R (x)$ is a disorder-dependent $SO(3)$ matrix with which the bare action together with Eq.~\eqref{eq:disorderAction} becomes the free-fermion action (see Ref.~\onlinecite{Levin2007} for details). Moreover, $L_x$ is the generator of $SO(3)$ describing rotation around the $x$-axis. Under the assumption that $\xi_{\rho \sigma, ab} \equiv v_{\rho \sigma}(R^T(x) L_x R(x))_{ab}$ is a Gaussian random variable, $\langle \xi_{\rho \sigma, ab}^{}(x)\xi_{\rho \sigma, a'b'}^*(x')\rangle =W_{\rho \sigma, ab} \delta(x-x') \delta_{a a'} \delta_{b b'}$, the disorder strengths $\tilde{W}_{\rho \sigma, ab}$  renormalize according to
\begin{align}
\label{eq:RG2LL}
d\tilde{W}_{\rho \sigma, ab} / d\ln  \ell = (3- 2\Delta_{\rho \sigma}) \tilde{W}_{\rho \sigma, ab} = - \tilde{W}_{\rho \sigma, ab}, 
\end{align}
since $\Delta_{\rho \sigma}=2$ (with respect to the disordered fixed point) in the absence of the interactions between the LLLs and the 2LL. When $T<\tilde{T}$, the RG flow in Eq.~\eqref{eq:RG2LL} terminates at the thermal length $L_T \propto 1/T$, where the disorder strengths are
\begin{align}
\label{eq:WatLT}
\tilde{W}_{\rho \sigma, ab} (L_T) = \tilde{W}_{\rho \sigma, ab}^0 \: \ell_0 / L_T.
\end{align}
Here, $\tilde{W}_{\rho \sigma, ab}^0 \equiv \tilde{W}_{\rho \sigma, ab} (\ell_0)$. Below we focus on $\tilde{W}_{\rho \sigma} \equiv \rm{max} \big[ \tilde{W}_{\rho \sigma, ab} \big]$ as it dominates in equilibrating the 2LL. Beyond $L_T$, $\tilde{W}_{\rho \sigma}$ scales classically, leading to 
\begin{align}
\label{eq:WatLeq}
\tilde{W}_{\rho \sigma} (L_T) / L_T = \tilde{W}_{\rho \sigma} (\ell_{\rm{eq}}^*) / \ell_{\rm{eq}}^*  \sim  1/ \ell_{\rm{eq}}^*,
\end{align}
where we defined $\ell_{\rm{eq}}^*$ as $ \tilde{W}_{\rho \sigma} (\ell_{\rm{eq}}^*) \sim 1$. Combining Eqs.~\eqref{eq:WatLT} and \eqref{eq:WatLeq}, we obtain the low-temperature scaling
\begin{align} 
\label{eq:leq*}
\ell_{\rm{eq}}^* \sim L_T^2 / \ell_0 \tilde{W}_{\rho \sigma}^0 \propto 1 /T^2,
\end{align}
in agreement with Ref.~\onlinecite{Ma2020}. For $T>\tilde{T}$ (see Fig.~\ref{fig:PhaseDiagram}), the RG flow terminates at $\ell = L_T$ before reaching the disorder fixed point. A similar RG analysis~\cite{SuppMat} results in the high temperature scaling $\ell_{\rm{eq}}^{*} \sim L_T (\ell_0/L_T)^{3-2\Delta_{\rm{2LL}}} \propto T^{2 - 2 \Delta_{\rm{2LL}}}$. The complete temperature scaling of $\ell_{\rm eq}^*$ is depicted in Fig.~\ref{fig:PhaseDiagram}. The scalings match at the crossover scale $L_T\sim \ell_0\Leftrightarrow T\sim \tilde{T}$. 
We now return to the vicinity of the fixed point, and consider weak random electron tunneling between the LLLs and the 2LL. The perturbing action reads
\begin{align}
S_{\rm{LLM}}&= \int dt dx e^{i \phi_1 (x) } e^{- 2i \phi_{\rho} (x)} \Big[\xi_{\rm{LLM, 1}} (x) \Big( \frac{\psi_2 - i\psi_3}{2} \Big) \nonumber \\
&+ \xi_{\rm{LLM, 2}} (x)\Big( \frac{\psi_2 + i\psi_3}{2} \Big) +
 \xi_{\rm{LLM, 3}} (x) \psi_1 + 
 \rm{H.c.} \Big], \notag
\end{align}
where $\psi_1\equiv \psi$, $\psi_2 = e^{i (\phi_3 + 2\phi_4) } + e^{-i (\phi_3 + 2\phi_4) } $, and $\psi_3 =  - i \left (e^{i (\phi_3 + 2\phi_4) } - e^{-i (\phi_3 + 2\phi_4) } \right )$. We neglect tunneling between $\phi_2$ and the 2LL, assuming negligible spin-flip tunneling. With respect to the fixed point, all tunneling operators have scaling dimensions $\Delta_{\rm{LLM}} = 2$. The disorder strengths are assumed Gaussian: $\langle \xi_{\rm{LLM}, i}^{} (x)\xi_{\rm{LLM}, i'}^*(x')\rangle =W_{\rm{LLM}, i} \delta(x-x') \delta_{i i'}$. The disorder strengths $\tilde{W}_{\rm{LLM}, i}$  then renormalize according to
\begin{align}
d\tilde{W}_{\rm{LLM}, i} / d\ln  \ell = (3- 2\Delta_{\rm{LLM}}) \tilde{W}_{\rm{LLM}, i} = - \tilde{W}_{\rm{LLM}, i}. 
\end{align}
Again, we consider only the dominating disorder $\tilde{W}_{\rm{LLM}} \equiv \textrm{max} \big[\tilde{W}_{\rm{LLM}, i} \big]$. Following the procedure leading to Eqs.~\eqref{eq:WatLT}-\eqref{eq:leq*}, we arrive at the length scale $\ell_{\rm eq}$, governing the LLM. It scales as 
\begin{align}
\label{eq:LowT2}
\ell_{\rm{eq}} \sim L_T^2 / \ell_0 \tilde{W}_{\rm{LLM}}^0 \propto 1 /T^2\,, 
\end{align}
where $\tilde{W}_{\rm{LLM}}^0$ is the disorder strength with the largest value at $\ell = \ell_0$. The low $T$ scaling of $\ell_{\rm eq}$ is depicted in Fig.~\ref{fig:PhaseDiagram}. Our results \eqref{eq:leq*} and~\eqref{eq:LowT2} imply that $\ell_{\rm{eq}}^* \ll \ell_{\rm{eq}}$ (at least for sufficiently low $T$) and thus
the transport regime $\mathcal{II}$ holds in a broad range of $T$.

\textit{Numerical computation of the noise.---}
We next turn to a computation of the noise scaling using the model from Refs.~\onlinecite{Park2019,Spanslatt2019}. We introduce a set of virtual reservoirs
attached to each channel along the edge. Such reservoirs define and maintain local equilibrium conditions in each channel~\cite{Nosiglia2018}. In the continuum limit, we obtain a set of transport equations for the local voltages, local temperatures, and the local noise along the edge~\cite{SuppMat}. By numerically solving these equations for the aPf edge, we obtain the plots in Fig.~\ref{fig:Fig1}. In regime $\mathcal{I}$, $S$ rises first linearly, and then drops exponentially in $L/\ell^*_{\rm eq}$. Around $\log[L/\ell^*_{\rm eq}]\approx 2$ (regime $\mathcal{II}$), $S\simeq c_1 - c_2 \sqrt{L/\ell_{\textrm{eq}}}$. The algebraic corrections to the constant scaling become suppressed for larger $\ell_{\rm eq}$ and develops into a plateau. On this plateau $G^Q/\kappa T\approx 5/2$. In regime $\mathcal{III}$, $S\simeq e^{-L/\ell^*_{\rm eq}}$ and $G^Q/\kappa T=3/2$. Figs.~\ref{fig:TAB} and~\ref{fig:TB} depict the edge channel temperature profiles in regimes $\mathcal{II}$ and $\mathcal{III}$ respectively. In the former regime, heat flows ballistically upstream with diffusive corrections from LLM. In the latter, the upstream heat propagation is exponentially suppressed in $L$.

\textit{Discussion.---}
We now justify the assumption of weak LLM, i.e., that typical experimental conditions favor regime $\mathcal{II}$. Since $\phi_1$ and the 2LL (having the same spin polarization) are spatially far apart, electron tunneling between these levels can be assumed to be weak. By contrast, $\phi_2$ and the 2LL are spatially closer, but have opposite spin-polarizations and tunneling between them is therefore also strongly suppressed, assuming no (or only weak) spin-rotation symmetry breaking. Strong inter-channel interactions may also weaken the LLM~\cite{Asasi2020}. Moreover, upstream heat propagation at $\nu=5/2$ was reported in Ref.~\onlinecite{Bid2010}, providing further support for regime $\mathcal{II}$.

Our proposed measurement of $S$ should be feasible with present technology. We envision a device capable of measuring both $G^Q$ and $S(L/\ell^*_{\rm eq})$. The latter measurement can be performed either by varying the inter-contact distance $L$, e.g., with a modulation gate, or by using several contacts spaced along the edge. Another possibility is to fix $L$ and instead tune the equilibration length, as recently was demonstrated in a specially designed double-well device at $\nu=2/3$~\cite{Cohen2019}. Our setup allows in principle for observing a transition of $G^Q/\kappa T$ from $5/2$ to $3/2$ with increasing $L$, which would strongly favor the aPf state (see Tab.~\ref{tab:TransportTab}). 
 
 Our analysis is based on no heat leakage into the bulk. If such leakage occurs, the relation between bulk topological order and edge heat transport breaks down~\cite{Aharon2019}. No-leakage experimental conditions were demonstrated at $\nu=5/2$ in Ref.~\onlinecite{Banerjee2018}. 

\textit{Summary.---}
We studied shot noise $S$ on the $\nu=5/2$ FQH edge for the three main edge candidates consistent with half-integer quantization of $G^Q$: Pfaffian, particle-hole Pfaffian, and anti-Pfaffian. Assuming full equilibration, which requires strong Landau level mixing, we argued that $S$ vanishes or decays exponentially in the edge length for all three candidates. However, in the regime where Landau level mixing is negligible, but intra-Landau level equilibration is efficient, only the anti-Pfaffian edge generates non-vanishing $S$. We demonstrated that a transport regime with $G^Q/\kappa T=5/2$ in combination with $S\simeq c_1 - c_2 \sqrt{L/\ell_{\textrm{eq}}}$ uniquely singles out the anti-Pfaffian. By contrast, for the same $G^Q$, the scaling $S\simeq 0$ points instead strongly towards the particle-hole Pfaffian. The Pfaffian edge exhibits robustly $G^Q / \kappa T = 7/2$ and $S=0$. We expect our results to be very useful for experimentally determining the $\nu=5/2$ edge structure. Our analysis can also be extended to other FQH states. 
\begin{acknowledgments}
\textit{Acknowledgments.---}
We thank A. Stern, Y. Oreg, B. Dutta, R. Melcer, and M. Heiblum for helpful discussions. C.S.,  Y.G., and  A.D.M. acknowledge support by DFG Grant No. MI 658/10-1 and by the German-Israeli Foundation Grant No. I-1505-303.10/2019. Y.G. further acknowledges support by DFG RO 2247/11-1, CRC 183 (project C01), and the Minerva foundation. J.P acknowledges support from CRC 183 (project A01). J.P. and C.S. contributed equally to this work.

\end{acknowledgments}


\begin{thebibliography}{67}
\bibitem [{Gam()}]{GammaNote}%
  \BibitemOpen
  \href@noop {} {\bibinfo  {journal} {We assume the heat and charge
  equilibration lengths to be equal. In principle, the heat equilibration
  length can be much larger than that for charge due to inter-channel
  interactions. In this case, our conclusions remain valid upon interpreting
  $\ell_{\rm eq}$ as the heat equilibration length}\ }\BibitemShut {NoStop}%
\bibitem [{LLL()}]{LLLEquili}%
  \BibitemOpen
  \href@noop {} {\bibinfo  {journal} {Equilibration between channels in the
  LLLs is immaterial since they both propagate downstream}\ }\BibitemShut
  {NoStop}%
\end{thebibliography}

%


\clearpage
\newpage
\onecolumngrid
\global\long\def\thesection{S\Alph{section}}
\global\long\def\thesubsection{\Roman{subsection}}
\setcounter{equation}{0}
\setcounter{figure}{0}
\setcounter{table}{0}
\setcounter{page}{1}
\renewcommand{\theequation}{S\arabic{equation}}
\renewcommand{\thefigure}{S\arabic{figure}}
\renewcommand{\bibnumfmt}[1]{[S#1]}
\renewcommand{\citenumfont}[1]{S#1}

\bigskip
\begin{center}
\large{\bf Supplemental Material for "Noise on the Non-Abelian $\nu=5/2$ Fractional Quantum Hall Edge"\\}
\end{center}
\begin{center}
Jinhong Park$^{1,2}$, Christian Sp\r{a}nsl\"{a}tt$^{3,4}$, Yuval Gefen,$^{2,3}$ and Alexander D. Mirlin$^{3,4,5,6}$
\\
{\it $^{1}$Institute for Theoretical Physics, University of Cologne, Z\"{u}lpicher Str. 77, 50937 K\"{o}ln, Germany \\$^{2}$Department of Condensed Matter Physics, Weizmann Institute of Science, Rehovot 76100, Israel\\$^{3}$Institute for Quantum Materials and Technologies, Karlsruhe Institute of Technology, 76021 Karlsruhe, Germany \\$^4$Institut f\"{u}r Theorie der Kondensierten Materie, Karlsruhe Institute of Technology, 76128 Karlsruhe, Germany \\ $^{5}$Petersburg Nuclear Physics Institute, 188300 St. Petersburg, Russia
\\ $^{6}$L.\,D.~Landau Institute for Theoretical Physics RAS, 119334 Moscow, Russia} \\
(Dated: \today)
\end{center}

In this Supplemental Material, we provide additional details in the derivation of the equilibration length scaling laws (Sec.~\ref{sec:RGDetails}), and details in our numerical calculations of the anti-Pfaffian noise characteristics (Sec.~\ref{sec:EquilibrationAndNoise}).
\section{Details of the RG analysis}
\label{sec:RGDetails}
Here, we provide additional details in the derivation of the scaling behavior of the anti-Pfaffian (aPf) equilibration lengths $\ell_{\rm eq}^*$ and $\ell_{\rm eq}$. To this end, we follow the renormalization group (RG) approach in Ref.~\onlinecite{Protopopov2017SM}. We start by focusing on the case of strong interactions ($\Delta_{\rm{2LL}} < 3/2$) within the \rm{2LL}. The edge of the aPf state are described by the action 
\begin{align}
S_0 + S_{\psi} = \int dx dt \left [- \sum_{ij} \frac{1}{4\pi} \left (K_{ij} \partial_x \phi_i \partial_t \phi_j + V_{ij} \partial_x \phi_i \partial_x \phi_j \right)  +  i \psi ( \partial_t - v_n \partial_x ) \psi \right],
\end{align} 
where $K = \rm{diag} (1, 1, 1, -2)$ in the basis of $(\phi_1, \phi_2, \phi_3, \phi_4)$. 
This action is integrable and thus it does not contain any mechanism for equilibration between the channels. Such a mechanism can however be captured by introducing random inter-channel tunneling. Assuming that the channels in the second Landau level (2LL) are spatially far away from the integer channels in the lowest Landau level (LLL), the equilibration dominantly occurs within in the 2LL. The random disorder term then reads~\cite{Levin2007SM}
\begin{equation} \label{eq:disorder}
	S_{2LL} =  \int dx dt \left [\xi_{\rm{2LL}}(x) \psi e^{i2\phi_4 + i\phi_3} + \rm{H.c.} \right].
\end{equation}
Here, $e^{i\phi_3}$ annihilates a right-moving electron while $\psi e^{i2\phi_4}$ creates a left-moving electron. For simplicity, we take $\xi_{\rm{2LL}}(x)$ as an uncorrelated complex Gaussian random variable satisfying $\langle \xi_{\rm{2LL}}^{} (x)\xi_{\rm{2LL}}^*(x')\rangle =W_{\rm{2LL}} \delta(x-x')$. We note that $\psi$ must be included in the left-moving electron operator to ensure the correct Fermionic commutation relation
\begin{align}
\left ( \psi (x) e^{i 2\phi_4 (x) } \right) \left ( \psi (x') e^{i 2\phi_4 (x') } \right) & = - \left ( \psi (x') e^{i 2\phi_4 (x') } \right)  \left ( \psi (x) e^{i 2\phi_4 (x) } \right)
e^{- 4[\phi_4 (x), \phi_4 (x') ]} \nonumber \\ &
= -\left ( \psi (x') e^{i 2\phi_4 (x') } \right) \left ( \psi (x) e^{i 2\phi_4 (x) } \right),
\end{align}
where we used the commutation relation $[\phi_4 (x), \phi_4 (x') ]=-i \pi \text{sgn}(x-x')/2$. 
Eq.~\eqref{eq:disorder} leads to the first order RG equation
\begin{align} \label{eq:RGequation}
\frac{d\tilde{W}_{\rm{2LL}}}{d\ln  \ell} = (3- 2\Delta_{\rm{2LL}}) \tilde{W}_{\rm{2LL}}, 
\end{align}
with the running length scale denoted $\ell$. Furthermore, $\Delta_{\rm{2LL}}$ is the scaling dimension of the electron tunneling operator in Eq.~\eqref{eq:disorder}, $\tilde{W}_{\rm{2LL}}$ the dimensionless disorder strength renormalizing $W_{\rm{2LL}}$; Hereafter we will denote all dimensionless disorder strengths with tildes. In the vanishing interaction limit ($V_{13}= V_{14} = V_{23} = V_{24} = 0$) between the lowest Landau level (LLL) and 2LL, $\Delta_{\rm{2LL}}$ is computed as
\begin{align}
\Delta_{\rm{2LL}} = \frac{1}{2} + \frac{(3/2 - 2 x) }{\sqrt{1 - 2 x^2}}.
\end{align}
where $x = 2V_{34}/ (2V_{33} + V_{44})$. Note that
the term $1/2$ in $\Delta_{\rm{2LL}}$ originates from the Majorana mode.
When the perturbation Eq.~\eqref{eq:disorder} is relevant ($\Delta_{\rm{2LL}} < 3/2$), disorder drives the system towards the $\Delta_{\rm{2LL}} = 1$ disordered fixed point~\cite{Levin2007SM}.
Eq.~\eqref{eq:RGequation} then defines an elastic length scale $\ell_0$ over which disorder mixes the channels within the 2LL. Let $\ell_0$ be the value of the running $\ell$ at which $\tilde{W}_{\rm{2LL}}$ becomes unity: 
\begin{align} \label{eq:elasticlength}
\ell_0 \sim a \tilde{W}_{\rm{2LL}, 0}^{1/(3-2\Delta_{\rm{2LL}})}, 
\end{align}
where $a$ is our ultra-violet length cutoff (e.g., the lattice constant) and $\tilde{W}_{\rm{2LL}, 0}$ is the bare disorder strength at $\ell = a$. 
As the system size $L$ goes beyond $\ell_0$, the system flows
to $\Delta_{\rm{2LL}} = 1$ with three upstream-propagating neutral Majorana modes $\psi_{a}$ ($a = 1,2,3$) and three downstream-propagating charge bosonic modes $\phi_1$, $\phi_2$, and $\phi_{\rho}$. After reaching the vicinity of the fixed point, $\ell_0$ acts as the new ultra-violet cutoff of the RG analysis. 

We now turn our attention to the scaling of the equilibration lengths in temperature. An applied voltage bias $V$ substitutes $T$ when $eV \gg k_B T$, but at the moment, let us focus on the case of $k_B T \gg eV$. We first consider sufficiently low temperature ($T < \tilde{T}$ in Fig.~\ref{fig:PhaseDiagram} of the main text) such that the system arrives at the disorder fixed point. Inter-channel interactions between the LLL and 2LL are assumed negligible. The part of the action leading to the equilibration length $\ell_{\rm{eq}}^{*}$ within the 2LL is written as
\begin{equation}
 S_{\psi \rho} = -  \frac{v_{\rho \sigma}}{2\pi} \sum_{a \neq b} \int dx dt  \partial_x \phi_\rho \psi_a  (R^T(x) L_x R(x))_{ab} \psi_b.
\end{equation}
Here $R (x)$ is a disorder-dependent $SO(3)$ matrix with which the bare action together with Eq.~\eqref{eq:disorder} becomes the free-fermion action (see Ref.~\onlinecite{Levin2007} for a detailed description). Moreover, $L_x$ is the generator of $SO(3)$ describing rotation around the $x$ axis. Under the assumption that $\xi_{\rho \sigma, ab} \equiv v_{\rho \sigma}(R^T(x) L_x R(x))_{ab}$ follows the Gaussian random distribution $\langle \xi_{\rho \sigma, ab}^{} (x)\xi_{\rho \sigma, a'b'}^*(x')\rangle =W_{\rho \sigma, ab} \delta(x-x') \delta_{a a'} \delta_{b b'}$, the dimensionless disorder strengths $\tilde{W}_{\rho \sigma, ab}$ (proportional to $W_{\rho \sigma, ab}$) renormalize according to the RG equation
\begin{align} \label{eq:rgequation2LL}
\frac{d\tilde{W}_{\rho \sigma, ab}}{d\ln  \ell} = (3- 2\Delta_{\rho \sigma}) \tilde{W}_{\rho \sigma, ab} = - \tilde{W}_{\rho \sigma, ab}. 
\end{align}
Here, the scaling dimension $\Delta_{\rho \sigma}=2$ in the absence of the interactions between the LLL and 2LL. The RG flow terminates at the thermal length $L_T \propto 1/T$, where the dimensionless disorder strengths read 
\begin{align} \label{eq:eqlengthlowtemp}
\tilde{W}_{\rho \sigma, ab} (L_T) = \tilde{W}_{\rho \sigma, ab}^0 \frac{\ell_0}{L_T},
\end{align}
where $\tilde{W}_{\rho \sigma, ab}^0 \equiv \tilde{W}_{\rho \sigma, ab} (\ell_0)$. We focus next only on the largest of the of dimensionless disorder strengths ($\tilde{W}_{\rho \sigma} \equiv \rm{max} \left [ \tilde{W}_{\rho \sigma, ab} \right ]$) as it dominates when determining the equilibration length. 
Beyond the scale $L_T$, the scaling of $\tilde{W}_{\rho \sigma}$ continues classically (i.e, $\tilde{W}_{\rho \sigma}$ grows linearly) leading to 
\begin{align} \label{eq:eqlengthlowtempcla}
\tilde{W}_{\rho \sigma} (L_T) / L_T = \tilde{W}_{\rho \sigma} (\ell_{\rm{eq}}^*) / \ell_{\rm{eq}}^*  \sim  1/ \ell_{\rm{eq}}^* .
\end{align}
Here, $\ell_{\rm{eq}}^*$ is defined as the length at which $ \tilde{W}_{\rho \sigma} (\ell_{\rm{eq}}^*) \sim 1$. Combining Eqs.~\eqref{eq:eqlengthlowtemp} and \eqref{eq:eqlengthlowtempcla}, we obtain 
\begin{align} \label{eq:equilibrationlength*}
\ell_{\rm{eq}}^* \sim \frac{L_T^2}{\ell_0 \tilde{W}_{\rho \sigma}^0} \propto \frac{1}{T^2}. 
\end{align}

Close to the disordered fixed point, the channels in the 2LL start to couple to channels in the LLL by disorder. The dominant term to describe the coupling originates from the electron tunneling term with smallest scaling dimension as
\begin{align}
\label{eq:SLLM}
S_{\rm{LLM}} &= \int dx dt \Bigg[\xi_{\rm{LLM, 1}} (x) e^{i \phi_1 (x) } e^{- i \phi_3 (x)}
+ \xi_{\rm{LLM, 2}} (x) e^{i \phi_1 (x) } e^{-  i \left ( 3\phi_3 (x) + 4\phi_4 (x) \right)} +\xi_{\rm{LLM, 3}} (x) e^{i \phi_1 (x) } e^{-  i \left ( 2\phi_3 (x) + 2\phi_4 (x) \right)}\psi  \notag\\ &+ \rm{H.c.} \Bigg] = \int\emph{dxdt} \left[ e^{i \phi_1 (x) } e^{- 2i \phi_{\rho} (x)} \left (
\xi_{\rm{LLM, 1}} (x) \left ( \frac{\psi_2 - i\psi_3}{2} \right)
+ \xi_{\rm{LLM, 2}} (x)\left ( \frac{\psi_2 + i\psi_3}{2} \right) +
 \xi_{\rm{LLM, 3}} (x) \psi_1 + 
 \rm{H.c.} \right ) \right].  
\end{align}
Here, $\phi_{\rho} = (\phi_3 + \phi_4 )$, $\psi_1= \psi$, $\psi_2 = e^{i (\phi_3 + 2\phi_4) } + e^{-i (\phi_3 + 2\phi_4) } $, and $\psi_3 =  - i \left (e^{i (\phi_3 + 2\phi_4) } - e^{-i (\phi_3 + 2\phi_4) } \right )$.
In the vicinity of the disordered fixed point, all terms have the same scaling dimension: $\Delta_{\rm{LLM}} = 2$ and the disorder strengths satisfy Gaussian distributions $\langle \xi_{\rm{LLM}, i}^{} (x)\xi_{\rm{LLM}, i'}^*(x')\rangle =W_{\rm{LLM}, i} \delta(x-x') \delta_{i i'}$. The dimensionless disorder strengths $\tilde{W}_{\rm{LLM}, i}$ (proportional to $W_{\rm{LLM}, i}$) renormalize according to
\begin{align}
\frac{d\tilde{W}_{LLM, i}}{d\ln  \ell} = (3- 2\Delta_{\rm{LLM}}) \tilde{W}_{\rm{LLM}, i} = - \tilde{W}_{\rm{LLM}, i}. 
\end{align}
We next only consider renormalization of the strongest disorder $\tilde{W}_{\rm{LLM}} \equiv \textrm{max} \left [\tilde{W}_{\rm LLM, i} \right ]$ as it dominates in the equilibration. Following the same procedure of Eqs.~\eqref{eq:rgequation2LL}-\eqref{eq:equilibrationlength*}, we arrive at 
\begin{align}
\ell_{\rm{eq}} \sim \frac{L_T^2}{\ell_0 \tilde{W}_{\rm{LLM}}^0} \propto \frac{1}{T^2}, 
\end{align}
where $\tilde{W}_{}^0$ is the disorder strength with the largest value at $\ell = \ell_0$. 
When the temperature is larger than $\tilde{T}$ (see Fig.~\ref{fig:PhaseDiagram} in the main text), on the other hand, Eq.~\eqref{eq:disorder} terminates before reaching the disorder fixed point. At the thermal length scale $\ell = L_T$, the dimensionless disorder strength $\tilde{W}_{\rm{2LL}}$ reads
\begin{align}
\tilde{W}_{\rm{2LL}} (L_T) =  \tilde{W}_{\rm{2LL}, 0} \left (\frac{L_T}{a} \right)^{3-2\Delta_{\rm{2LL}}} = \left (\frac{L_T}{\ell_0} \right)^{3-2\Delta_{\rm{2LL}}}.
\end{align}
Following the same procedure as Eqs.~\eqref{eq:eqlengthlowtempcla} and~\eqref{eq:equilibrationlength*} beyond the thermal length scale $\ell = L_T$, we obtain 
\begin{align}
\ell_{\rm{eq}}^{*} = L_T \frac{\tilde{W}_{\rm{2LL}} (\ell_{\rm{eq}}^{*})}{\tilde{W}_{\rm{2LL}}  (L_T)} \sim L_T \left (\frac{\ell_0}{L_T} \right )^{3-2\Delta_{\rm{2LL}}} \propto T^{2 - 2 \Delta_{\rm{2LL}}}.
\end{align}
The scaling of $\ell_{\rm{eq}}$ at high temperature ($T > \tilde{T}$)  is much more complicated to analyze, since it generally depends on the scaling dimensions and the bare strengths of many possible disorder terms. However, under the assumption that $W^0_{\rm LLM,i}$ is strongly dominating, we find that $\ell_{\rm eq} \sim T^{2-2\Delta_{\rm LLM,i}}$, where $\Delta_{\rm{LLM},i}$ is the scaling dimension of the operator associated to $\xi_{\rm{LLM},i}$. 

\section{Numerical calculation of equilibration and noise for the aPf edge}
\label{sec:EquilibrationAndNoise}
To model equilibration and noise, we use the theory developed in Refs.~\onlinecite{Park2019SM,Spanslatt2019SM}. We denote with $\vec{V}(x)=(V_{12},V_3,V_4)^T(x)$ (superscript $T$ denotes transposition) the local voltages of the bosonic channels $\phi_1+\phi_2$ (assumed for simplicity to be in equilibrium upon exiting the contacts), $\phi_3$, and $\phi_4$ respectively. The voltages evolve along an aPf edge according to the transport equation
\begin{equation}
\label{eq:VoltageEquation}
\partial_{x}\vec{V}(x) = \mathcal{M}_V \vec{V}(x), \quad 
\mathcal{M}_V = \frac{1}{\ell^*_{\rm eq}}\begin{pmatrix}
-\alpha & \frac{\alpha}{2} & \frac{\alpha}{2}\\
\alpha & -1-\alpha & 1 \\
-2\alpha & -2 & 2\alpha + 2
\end{pmatrix},
\end{equation}
where $\alpha\equiv\ell^*_{\rm eq}/\ell_{\rm eq}\ll 1$ is a parameter determining the degree of Landau level mixing. For simplicity, we ignore any temperature or voltage dependence of the equilibration lengths which we choose as constant. The corresponding local electrical currents $\vec{I}(x)$ obey a similar equation
\begin{equation}
\label{eq:CurrentEquation}
\partial_{x}\vec{I}(x) = \mathcal{M}_I \vec{I}(x), \qquad \mathcal{M}_I =   \mathcal{D} \mathcal{M}_V  \mathcal{D}^{-1},
\end{equation}
 with  $\mathcal{D} = \text{diag}(2,1,-1/2)$. Note that the charge neutral Majorana mode is absent in determining the voltage and current profiles. The local temperatures are governed by
\begin{equation}
\label{eq:TemperatureEquation}
\partial_x\vec{T^2}(x) = \mathcal{M}_T \vec{T^2}(x) + \Delta\vec{V}(x), \quad \mathcal{M}_T =
\frac{1}{\ell^*_{\rm eq}}\begin{pmatrix}
-\alpha & \frac{\alpha}{2} & \frac{\alpha}{2}\\
\alpha & -1-\alpha & 1 \\
-\frac{2}{3}\alpha & -\frac{2}{3} & \frac{2\alpha + 2}{3}
\end{pmatrix},
\end{equation}
with $\vec{T^2}(x)=(T_{12}^2,T_3^2,T_4^2)^T(x)$, and we have for simplicity assumed that the charge and heat equilibration lengths are identical.

 Generally, these two length scales can differ depending on the microscopic details of the edge, but this complication does not change our qualitative results. Moreover,
\begin{equation}
\label{eq:JouleHeating}
\Delta \vec{V}(x) =\frac{e^2}{h\kappa} \left(\alpha\frac{(V_{12}-V_4)^2}{2\ell^*_{\rm eq}}+\alpha\frac{(V_{12}-V_3)^2}{2\ell^*_{\rm eq}}, \frac{(V_{3}-V_4)^2}{\ell^*_{\rm eq}}+\alpha\frac{(V_{12}-V_3)^2}{\ell^*_{\rm eq}},-3\alpha\frac{(V_{12}-V_4)^2}{2\ell^*_{\rm eq}}-3\frac{(V_{3}-V_4)^2}{2\ell^*_{\rm eq}} \right)^T(x)
\end{equation}
reflects the Joule heating contribution. In contrast to Eq.~\eqref{eq:VoltageEquation}, the Majorana mode contributes in Eq.~\eqref{eq:TemperatureEquation} by the fractional pre-factor $2/3$ (the inverse central charge of the 2LL left-movers). Since only the combined operator $\psi e^{2i\phi_4}$ constitutes an electron, we assume for simplicity that $\psi$ and $\phi_4$ are always thermally equilibrated and their common temperature is $T_4(x)$. 

Shot noise is related to the local current fluctuations, which are found from
\begin{equation}
\label{eq:NoiseEquation}
\partial_x\vec{\delta I^{}}(x) = \mathcal{M}_I \vec{\delta I^{}}(x) + \vec{\delta I}^{\tau,{\rm int}}(x), \quad \vec{\delta I}^{\tau,{\rm int}}(x) = \begin{pmatrix}
	-1 & -1 & 0 \\
	1 & 0 & -1 \\
	0 & 1 & 1
\end{pmatrix}
\begin{pmatrix}
	\delta I^{\tau,{\rm int}}_{12,3}(x) \\
	\delta I^{\tau,{\rm int}}_{12,4}(x) \\
	\delta I^{\tau,{\rm int}}_{3,4}(x)
\end{pmatrix},
\end{equation}
in which $\delta I_{n=12,3,4}^{\tau,{\rm int}}(x)$ are intrinsic fluctuations governed by local equilibrium noise relations
\begin{subequations}
\label{eq:NoiseCorrelator}
\begin{align}
\overline{\delta I^{\tau,{\rm int}}_{12,3}(x)\delta I^{\tau,{\rm int}}_{12,3}(y)} &= \frac{2e^2}{h\ell_{\rm eq }}k_B\left[T_{12}(x)+T_{3}(y)\right]\delta(x-y), \\
 \overline{\delta I^{\tau,{\rm int}}_{12,4}(x)\delta I^{\tau,{\rm int}}_{12,4}(y)} &= \frac{2e^2}{h\ell_{\rm eq }}k_B\left[T_{12}(x)+T_{4}(y)\right]\delta(x-y), \\
 \overline{\delta I^{\tau,{\rm int}}_{3,4}(x)\delta I^{\tau,{\rm int}}_{3,4}(y)} &= \frac{2e^2}{h\ell^*_{\rm eq }}k_B\left[T_{3}(x)+T_{4}(y)\right]\delta(x-y),
\end{align}
\end{subequations}
all other correlators being zero. The overline means time-average. Finally, the noise in any of the two contacts on the edge (equal due to current conservation) is formally defined by
\begin{align}
\label{eq:NoiseFormal}
S = \overline{\left(\delta I_{12}(L)+\delta I_3(L)\right)^2} = \overline{\left(\delta I_4(0)\right)^2}.
\end{align}
To compute $S$, we first choose boundary conditions $V_{12}(0)=V_3(0)=V_0$, $V_4(L)=0$, which by Eq.~\eqref{eq:VoltageEquation} gives the distribution of voltages in the bosonic channels. Next, we solve Eq.~\eqref{eq:TemperatureEquation} with boundary conditions $T_{12}(0)=T_3(0)=T_4(L)=0$ using the solutions $\vec{V}(x)$ in the Joule heating contribution~\eqref{eq:JouleHeating}. In our setup, we assume no heat bias during the noise measurement. Finally, with the obtained temperature profiles, we compute the average in Eq.~\eqref{eq:NoiseFormal} using the relations in Eq.~\eqref{eq:NoiseCorrelator}. If one is interested only in complete edge equilibration, one may reduce the edge into two counter-propagating hydrodynamic modes~\cite{Spanslatt2019SM} and the noise can be solved for analytically. To fully capture the transition between transport regimes $\mathcal{II}$ and $\mathcal{III}$, we have to resort to numerical calculations. 

In Fig.~\ref{fig:Noise} of the main text, we computed $S(L/\ell^*_{\rm eq})$ from $L=0.007$ to $L\approx 241$ with $\ell^*_{\rm eq}=1$ and $\alpha=0.01$ (blue solid line) and $\alpha=0.001$ (dashed blue line). We also used Eq.~\eqref{eq:TemperatureEquation} with $\Delta\vec{V}(x)=0$ to compute the thermal conductance $G^Q/(\kappa T)[L/\ell^*_{\rm eq}]$, where $T$ is the temperature bias between $x=0$ and $x=L$. Furthermore, Figs.~\ref{fig:TAB} and~\ref{fig:TB} depict solutions to Eq.~\eqref{eq:TemperatureEquation} for $L\approx 11.8$ and $L\approx 241$ respectively. For these two plots, we have taken $\ell^*_{\rm eq}=1$, $\alpha=0.01$.

\end{document}